\documentclass[prb,aps,amsmath,amssymb,reprint,showpacs,floatfix,superscriptaddress,longbibliography]{revtex4-1}
\usepackage{graphicx}% Include figure files
\usepackage{dcolumn}% Align table columns on decimal point
\usepackage{bm}% bold math
\usepackage{mathtools}% math tools
\usepackage[dvipsnames]{xcolor}
\begin{document}

%\preprint{Test}

%\title{Rashba mixing and orbital motion effects into transition phase laws in 
%2D Majorana nanowires} %Title of paper

\title{Current distributions in stripe Majorana junctions}


\author{Javier Osca}
\email{javier@ifisc.uib-csic.es}
\affiliation{Institut de F\'{\i}sica Interdisciplin\`aria i de Sistemes Complexos IFISC (CSIC-UIB), E-07122 Palma de Mallorca, Spain}
\author{Lloren\c{c} Serra}
%\email{llorens.serra@iuib.es}
\affiliation{Institut de F\'{\i}sica Interdisciplin\`aria i de Sistemes Complexos IFISC (CSIC-UIB), E-07122 Palma de Mallorca, Spain}
\affiliation{Departament de F\'{\i}sica, Universitat de les Illes Balears, E-07122 Palma de Mallorca, Spain}

\date{June 30, 2016}

\begin{abstract}
We calculate current and density distributions in stripe (2D planar) junctions between normal and Majorana 
nanowires having a finite ($y$)  transverse length.  
In presence of a magnetic field with vertical and in-plane components, the $y$-symmetry of the charge current distribution in the normal lead changes strongly across the Majorana phase transition:
from center-symmetric 
if a Majorana mode is present to laterally-shifted (as expected by the Hall effect) if the field is tilted such as to destroy the Majorana mode due to the projection rule. We compare quasi-particle and charge distributions of current and density, as well as spin magnetizations. The Majorana mode causes opposite spin accumulations on the junction
and the emergence of a spin current. 
\end{abstract}

\pacs{73.63.Nm,74.45.+c}% insert suggested PACS numbers in braces on next line

\maketitle %\maketitle must follow title, authors, abstract and \pacs

%\tableofcontents
%\section{Introduction}
%\label{intro}

\section{Introduction}

%\red{Comments:
%Topology: in a topological conductor the currents locate in parts of the system;
%Kwant: leads with SO and magnetic field are not supported in Kwant?
%High resolution of our matching method, no need of large grid?
%Our method is equivalent but superior to matching?
%}

\label{intro}

Majorana modes appearing at the ends of hybrid semiconductor-superconductor 
nanowires have been the focus of a strong interest in recent 
years.\cite{Alicea,Beenakker,StanescuREV,Franz}
The existence of these intriguing states was predicted theoretically \cite{Fu,Lutchyn,Oreg} and,
soon after, strong
experimental evidences were seen in a first round of experiments measuring the zero-bias anomaly associated with such modes.\cite{Mourik,Deng,Das,Finck} 
Recently, more refined experiments heavily suppressing disorder due to impurities and imperfections have 
confirmed the existence of these rigid and stable zero modes
showing an improved agreement with the theoretical expectations.\cite{Zhang}

The physics of hybrid Majorana nanowires was initially addressed with purely 1D models that nicely 
captured the basic behaviors. Two conditions need to be fulfilled for the existence of a Majorana 
mode. Namely, a critical field rule,  $\Delta_B > \sqrt{\Delta_0^2+\mu^2}$; and 
a field projection rule,
$\Delta_B \sin\theta\, |\sin\phi| < \Delta_0$, 
where $\Delta_B$ is the Zeeman energy associated with the modulus of the magnetic field
(but not with its direction),
$\Delta_0$ is the superconductor gap, $\mu$ is the wire chemical potential                 
while $\theta$ and $\phi$ are the polar and azimuthal  angles, respectively, 
of the magnetic field
assuming the 1D wire is on the Cartessian $x$ axis.
Extensions to take into account transverse motion were soon investigated considering a few coupled modes in multimode wires (quasi 1D models).\cite{Lutchyn2,Lutchyn3,Stanescu,Potter,Potter2,Law11,Lutchyn2,Tew12,sanjose,Wan15} 
A natural question appearing when including
higher dimensions is the role of orbital effects of the magnetic field. They have been studied in continuum 
models, numerically solved using grid discretizations, for the cases of planar stripes, cylindrical shells
and 3D hexagonal wires.\cite{Wim11,Lim3,Os14,Os15,Nij16,Sed16}

In this work we consider a 2D stripe of width $L_y$ containing a junction between a normal and a superconducting (Majorana) section. Transport is along $x$ and we aim specifically at discussing 
current and density spatial distributions and their dependence on the topological state of the 
Majorana nanowire. A general characteristic of topological materials is the distribution of currents on the system borders or edges.  
Ours is a minimal continuum model beyond strict 1D and it can be implemented by laterally patterning a 2D electron gas. Indeed, the proximity coupling of a 2D semiconductor heterostructure  and a superconductor
has already  been achieved in Refs.\ \onlinecite{Sha15,Kja16}, where a quantum point contact 
has been shown to present  a hard gap behavior and the conductance quantization typical of
normal-superconductor junctions.\cite{Bee92}
This geometry differs from the core-shell structure of Refs.\ \onlinecite{Mourik,Deng,Das,Finck}, but it is of interest as it provides relevant physical insights and may yield new technology opportunities.

Incident electrons from the normal side of the stripe can be reflected as holes by means of Andreev reflection. In presence of a Majorana mode this process dominates, even if a moderate potential barrier is present 
at the junction. Electron and hole quasiparticle currents flow in opposite directions on the stripe sides
but, since they yield charge currents in the same direction,  a symmetric transverse distribution of current
in the normal lead is found due to the Majorana mode. If, however, the Majorana mode is 
absent then Andreev reflection is suppressed, currents are purely of electron type and they yield
an asymmetric transverse profile on the stripe in a vertical magnetic field. 
We show below that this scenario is indeed found
in stripe Majorana junctions. 

By simply varying the orientation of the magnetic field, while keeping all
other parameters fixed, the Majorana state may be switched on and off, with the corresponding
modifications of the transverse current distribution in the normal lead.
We also find that the spin distributions in presence of a Majorana mode are reminiscent of the spin Hall effect, with up and down spin concentrations on opposite sides of the stripe in the region of the Majorana mode. 
The appearance of spin-flip density distributions is a manifestation of the effective spin-triplet pairing in the Majorana nanowire.
This spin distribution is accompanied by the emergence of a spin current in the normal lead. 

A related previous work by us was presented in Ref.\ \onlinecite{Os15b}. There
we discussed the quasiparticle density and current distributions at the 
ends of a finite (closed) stripe.
The finite stripe is easier modeled 
theoretically, however, it does not 
allow any charge current flow as it may occur in the open junction 
of the present work. 
Open Majorana systems were also recently addressed in Ref.\ \onlinecite{Sed16},
although not focussing on spatial distributions of currents
but on the phase diagrams with varying transverse lengths.

Numerical modeling of Majorana systems attached to leads 
can be performed using software packages such as {\tt KWANT}\cite{Gro14} or 
{\tt MATHQ}.\cite{mathq} In these approaches emphasis is normally placed in allowing
greater
flexibility with the modeling of the central part of the system, to which
simple normal leads are attached. In our case, however, the situation is 
reversed since we consider more
involved leads 
having spin-orbit coupling, magnetic Zeeman and orbital effects,  as well as superconductivity (in the right lead),
while the junction itself is a simple potential
barrier. For this reason we have developed a direct junction solver 
using the lead modes calculated independently.
Specific advantages of our method are high computational efficiency, high spatial 
resolution in $y$ and arbitrary resolution in $x$ without computational cost.
The work is organized as follows.
Section II presents the model and used resolution techniques.
Section III discusses the results and Sec.\ IV finally summarizes and concludes the 
work.

\section{Model and resolution}

We consider a 2D stripe containing a junction between a normal and a superconducting 
lead. A Rashba spin-orbit coupling and a magnetic field are acting in all parts of the system.
Electrons are incident from the asymptotic part of the normal side and they are scattered, 
either transmitted or reflected, 
by the junction that may contain a potential barrier.

\subsection{Hamiltonian}

The system is described with
a Bogoliubov-deGennes Hamiltonian, similarly as  
in Ref.\ \onlinecite{Os15},
\begin{equation}
\label{E1}
\mathcal{H}_{\it BdG} 
= 
\mathcal{H}_0
+
\mathcal{H}_{sc}
+
\mathcal{H}_Z
+
\mathcal{H}_R
+
\mathcal{H}_{\it orb}\; .
\end{equation}
Specifically, the first contribution to Eq.\ (\ref{E1}) contains the kinetic and 
potential energies not depending on the magnetic field
\begin{equation}
\label{E1_0}
	\mathcal{H}_{0} = 
	\left( \frac{p_{x}^{2}+p_y^2}{2m} + V(x,y) -\mu \right)\tau_z\; ,
\end{equation}
where the potential $V(x,y)$ contains the transverse confinement due to an infinite square well
of length $L_y$, as well as finite potential barrier along $x$ centered at the junction position $x=0$, of height $V_b$ and length $L_b$.

The superconducting term in Eq.\ (\ref{E1}) is
\begin{equation}
\label{E1_sc}
	\mathcal{H}_{sc} = \Delta_0(x)\,\tau_x\; ,
\end{equation}
where $\Delta_0(x)$ represents the induced superconductivity parameter that
vanishes for $x<0$ and takes a constant value $\Delta_0$ for $x>0$.
The subsequent Zeeman and Rashba terms are
\begin{eqnarray}
\label{E1_Z}
	\mathcal{H}_{\it Z} &=& 
\Delta_B\, \left(\sin\theta \cos\phi\, \sigma_x	+ \sin\theta \sin\phi\, \sigma_y +\cos\theta\, \sigma_z \right ) \, ;\\
\label{E1_R}
	\mathcal{H}_{\it R} &=& 
\frac{\alpha}{\hbar}\, \left(\, p_x \sigma_{y} - p_y \sigma_{x}\, \right)\tau_z \, .
\end{eqnarray}
The last contribution to Eq.\ (\ref{E1}) contains the magnetic orbital terms
\begin{equation}
\label{E1_orb}
	\mathcal{H}_{\it orb} = 
	\frac{\hbar^2}{2 m l^4_z}y^2 \tau_z - \frac{\hbar^2}{m l^2_z}y p_x - \frac{\alpha}{l^2_z}y \sigma_y \;,
\end{equation}
where  $l_z$ is the magnetic length depending on the vertical component of the magnetic field
$l_z^2\equiv\hbar c/eB_z$.
The Nambu spinor convention in 
Eqs.\ (\ref{E1_0})-(\ref{E1_orb}), relating discrete components with spin $(\uparrow\downarrow)$
and isospin $(\Uparrow\Downarrow)$ is
$\Psi\equiv(
\Psi_{\uparrow\Uparrow},\Psi_{\downarrow\Uparrow},\Psi_{\downarrow\Downarrow},-\Psi_{\uparrow\Downarrow })^{T}$.

\subsection{Algorithm}

We are interested in finding solutions of Schr\"odinger's equation for a given energy $E$
\begin{equation}
\label{E7}
\left({\cal H}_{\it BdG}-E\right) \Psi(xy\eta_\sigma\eta_\tau)= 0\; ,
\end{equation}
where $\eta_\sigma$ and $\eta_\tau$ represent the discrete spin and isospin variables, respectively.
Our method relies on $k$-expansions of the 
wave function, including the possibility of complex $k$'s,\cite{Serra}
for the asymptotic normal (left L) and superconductor (right R) sides, combined with a grid discretization 
in the junction (center C) region. The algorithm is based on the quantum-transmitting-boundary method and, in practice, it amounts to an effective way of matching the two different sets of asymptotic solutions in 2D. Our method is devised to allow a high spatial resolution, made possible because only a relatively small number 
of grid points along $x$ is required, which allows using a large number of points along $y$. In addition,
the use of asymptotic $k$-modes allows extending the solutions an arbitrary distance into the leads.  

In a
generic contact $c=L,R$ the wave function can be expanded as
\begin{equation}
\label{eq8}
\Psi(xy\eta_\sigma\eta_\tau) =
\sum_{\alpha n_\alpha}
{d_{n_\alpha}^{(\alpha,c)}
\exp{\left[ik_{n_\alpha}^{(\alpha,c)}x\right]}
\phi_{n_\alpha}^{(\alpha,c)}(y\eta_\sigma\eta_\tau)
}\; ,
\end{equation}
where $\alpha=i,o$ referes generically to both input and output modes. The bookkeeping of modes 
requires using three labels, $(\alpha,n_\alpha,c)$, corresponding to mode type, 
mode number and contact, respectively. The set of complex amplitudes $\{d_{n_\alpha}^{(\alpha,c)}\}$
fully characterizes the asymptotic solution in contact $c$. At this  point, we assume that the wire mode
wave numbers and wave functions are known for a sufficiently large set, $\{
k_{n_\alpha}^{(\alpha,c)},\, \phi_{n_\alpha}^{(\alpha,c)}(y\eta_\sigma\eta_\tau) 
\}$.

We introduce a uniform grid of points, containing the junction ($C$)  and small portions of the leads ($L,R$). Our sought-after unknowns are the values of the wave function on the grid points as well as the set of output amplitudes 
$\{\Psi(xy\eta_\sigma\eta_\tau),d_{n_o}^{(o,L/R)}\}$.
Notice that input amplitudes must be supplied. The closed-system of linear equation reads
\begin{widetext}
\begin{eqnarray}
\label{E9}
\left({\cal H}_{\it BdG}-E\right) \Psi(xy\eta_\sigma\eta_\tau) &=& 0\; , 
\qquad\qquad\qquad\qquad\qquad\qquad\qquad\quad\hspace*{0.15cm}
 (xy)\in C \; ,\\
\label{E10}
\Psi(xy\eta_\sigma\eta_\tau) 
-
\sum_{n_o}
{d_{n_o}^{(o,c)}
\exp{\left[ik_{n_o}^{(o,c)}x\right]}
\phi_{n_o}^{(o,c)}(y\eta_\sigma\eta_\tau)
}
&=& 
\sum_{n_i}
{d_{n_i}^{(i,c)}
\exp{\left[ik_{n_i}^{(i,c)}x\right]}
\phi_{n_i}^{(i,c)}(y\eta_\sigma\eta_\tau)
}\; , 
\quad (xy,c)\in L/R \; ,\\
\label{E11}
\sum_{\eta_\sigma\eta_\tau}\int{dy\,
\phi_{m_o}^{(o,c)}(y\eta_\sigma\eta_\tau)^*
\, \Psi(x_cy\eta_\sigma\eta_\tau)} 
&-&
\sum_{n_o}
{d_{n_o}^{(o,c)}
\exp{\left[ik_{n_o}^{(o,c)}x_c\right]}
{\cal M}_{m_on_o}^{(oc,oc)}
}
=  \nonumber\\
 && \sum_{n_i}
{d_{n_i}^{(i,c)}
\exp{\left[ik_{n_i}^{(i,c)}x_c\right]}
{\cal M}_{m_on_i}^{(oc,ic)}
} \; ,
\qquad\hspace*{0.3cm}
 c\in L/R \; ,
\end{eqnarray}
\end{widetext}
where $x_c$ indicates the $x$ coordinate of the grid points on region $C$ at the boundary with contact $c=L/R$ and we defined the overlap matrices
\begin{equation}
{\cal M}_{m_\alpha n_\beta}^{(\alpha c,\beta c)}
=
\sum_{\eta_\sigma \eta_\tau}
\int{dy\,
\phi_{m_\alpha}^{(\alpha,c)}(y\eta_\sigma\eta_\tau)^*
\phi_{m_\beta}^{(\beta,c)}(y\eta_\sigma\eta_\tau)
}\; .
\end{equation}

The linear system posed by Eqs.\ (\ref{E9})-(\ref{E11}) is efficiently solved using sparse matrix 
routines\cite{Harwell}
once a particular non-vanishing input amplitude is assumed, e.g., $d_{n_i}^{(i,L)}=1$. Indeed, quite high spatial resolutions of 
current and density can be achieved as shown below.
Once the set of output amplitudes $\{d_{n_o}^{(o,c)}\}$
is known, the wave function can be arbitrarily extended into the asymptotic regions, as anticipated, since there the $x$ dependence is analytical, Eq.\ (\ref{eq8}).

\subsection{Complex k's}

The above method requires the knowledge of mode wave numbers and wave functions for each lead $\{k,\phi(y\eta_\sigma\eta_\tau)\}$, where we dropped for simplicity all mode labels. They can be obtained in a very efficient way with the diagonalization of a sparse non-Hermitian matrix. Notice that 
the wave number $k$, not the energy $E$, is the required eigenvalue and that the original 
Schr\"odinger problem is nonlinear (quadratic)  in $k$. A clever trick  allows a transformation   
into a linear eigenvalue problem by enlarging the space of wave function components.\cite{tis01,xie16}

Replacing $p_x\to\hbar k$ in the Hamiltonian we obtain (properly defining ${\cal H}_{A/B} $)
\begin{equation}
{\cal H}_{\it BDG} \to
{\cal H}_{A}+
 {\cal H}_{B}\, \ell_0 k
+ \frac{\hbar^2 k^2}{2m}\tau_z \; ,
\end{equation} 
where $\ell_0$ is a length unit that will be specified below. Define now $\phi_{s_\sigma s_\tau}(y)$, the  
spin-isospin components with
proper spinors $\chi_{s}$, as
\begin{equation}
\phi(y\eta_\sigma\eta_\tau)=
\sum_{s_\sigma s_\tau}{
\phi_{s_\sigma s_\tau}(y)
\,\chi_{s_\sigma}(\eta_\sigma)
\,\chi_{s_\tau}(\eta_\tau)
}\; ,
\end{equation}
and the corresponding 'enlarged' set of components 
\begin{equation}
\tilde\phi_{s_\sigma s_\tau}(y)
\equiv \ell_0 s_\tau\,k\,
\phi_{s_\sigma s_\tau}(y).
\end{equation}

With the above definitions we recast Schr\"odinger's eigenvalue problem into
the doubled system of equations
\begin{widetext}
\begin{eqnarray}
\label{E16}
 s_\tau\,
\tilde\phi_{s_\sigma s_\tau}(y) &=&
k\,\ell_0\, \phi_{s_\sigma s_\tau}(y)\; ,\\
\label{E17}
-\frac{2m\ell_0^2}{\hbar^2}
\sum_{y's'_\sigma s'_\tau}{
\left[
\rule{0cm}{0.5cm}
\langle y s_\sigma s_\tau | {\cal H}_A | y' s'_\sigma s'_\tau\rangle\,
\phi_{s'_\sigma s'_\tau}(y')
\right.}
\qquad\qquad
&&\nonumber\\
+
\left.
\rule{0cm}{0.5cm}
\langle
y s_\sigma s_\tau | {\cal H}_B | y' s'_\sigma s'_\tau
\rangle\,
s'_\tau\,
\tilde\phi_{s'_\sigma s'_\tau}(y')
\right]
+\frac{2m\ell_0^2}{\hbar^2} E\,
\phi_{s_\sigma s_\tau}(y)
&=& k\,\ell_0\, \tilde\phi_{s_\sigma s_\tau}(y)\; ,
\end{eqnarray}
\end{widetext}
that,  indeed, yields the wavenumber $k$ as eigenvalue.
While the starting $E$-eigenvalue problem, Eq.\ (\ref{E7}), is Hermitian
and yields real energy eigenvalues, the transformed $k$-eigenvalue system
is non Hermitian. In this case, this is a nice property since wave numbers
have indeed to be complex for evanescent modes. 
We have diagonalized the eigenvalue problem posed by Eqs.\ (\ref{E16}) and (\ref{E17}) with the 
Arpack library,\cite{arpack} obtaining the wave numbers ordered by increasing distance (in the complex plane)
from a given reference value.  We typically include in our calculation  
the $\approx 50$ modes closer to the origin.

\subsection{Spatial distributions}

We are interested in the spatial distributions of density and currents. The 
quasiparticle probability distributions $\rho_{qp}(x,y)$ and $\vec\jmath_{qp}(x,y)$ for the finite (closed)
stripe were already discussed, e.g., in Ref.\ \onlinecite{Os15b}. They fulfill the continuity equation
\begin{equation}
\label{E18}
\frac{\partial \rho_{qp}}{\partial t} = -\nabla \cdot\vec\jmath_{qp}\; ,
\end{equation}
and they are given in terms of the four-component wave function spinors $\Psi$ by
\begin{eqnarray}
\label{E19}
\rho_{qp}(x,y) &=& \Psi^*(x,y)\,\Psi(x,y)\; , \\
\label{E20}
\vec{\jmath}_{qp}(x,y) &=& \Re\left[\, \Psi^*(x,y)\, \vec{\hat v}_{qp}\, \Psi(x,y)\, \right]\;,
\end{eqnarray}
where we defined the vector operator
\begin{eqnarray}
\label{E21}
\vec{\hat v}_{qp} =
-i\frac{\hbar}{m}\nabla\, \tau_z &+& \frac{e}{mc}\vec{A} 
+ \frac{\alpha}{\hbar}\left(
\sigma_y\vec{u}_x
-\sigma_x \vec{u}_y \right) \tau_z\; .
\end{eqnarray}
Expressions for the charge  $\rho_c(x,y)$ and current $\vec\jmath_c(x,y)$ 
densities
are easily obtained form the corresponding quasiparticle distributions simply inserting 
an additional $-e\tau_z$ factor in Eqs.\ (\ref{E19}) and (\ref{E20}) 
\begin{eqnarray}
\label{E22}
\rho_{c}(x,y) &=& -e\, \Psi^*(x,y)\,\tau_z\,\Psi(x,y)\; , \\
\label{E23}
\vec{\jmath}_{c}(x,y) &=& -e\,\Re\left[\, \Psi^*(x,y)\, \vec{\hat v}_{qp}\tau_z\, \Psi(x,y)\, \right]\;.
\end{eqnarray}
Similar distributions for spin magnetization density $\rho_{sp}(x,y)$ and current 
$\vec{\jmath}_{sp}(x,y)$
read
\begin{eqnarray}
\rho_{sp}(x,y) &=& \Psi^*(x,y)\,\sigma_z\,\Psi(x,y)\; ,\\
\vec{\jmath}_{sp}(x,y) &=& \Re\left[\, \Psi^*(x,y)\, \vec{\hat v}_{qp}\sigma_z\, \Psi(x,y)\, \right]\;.
\end{eqnarray}
It is worth stressing that neither $\vec{\jmath}_{c}$ nor $\vec{\jmath}_{sp}$ fulfill a continuity equation similar to Eq.\ (\ref{E18}), since they are not conserved quantities in the sense that the 
superconductor may act as a source of charge and spin currents. Indeed, explicit examples where this occurs are shown in next section on results.

\begin{figure}[t]
\centering\resizebox{0.475\textwidth}{!}{
	\includegraphics[trim={0 0 0 0},clip=true]{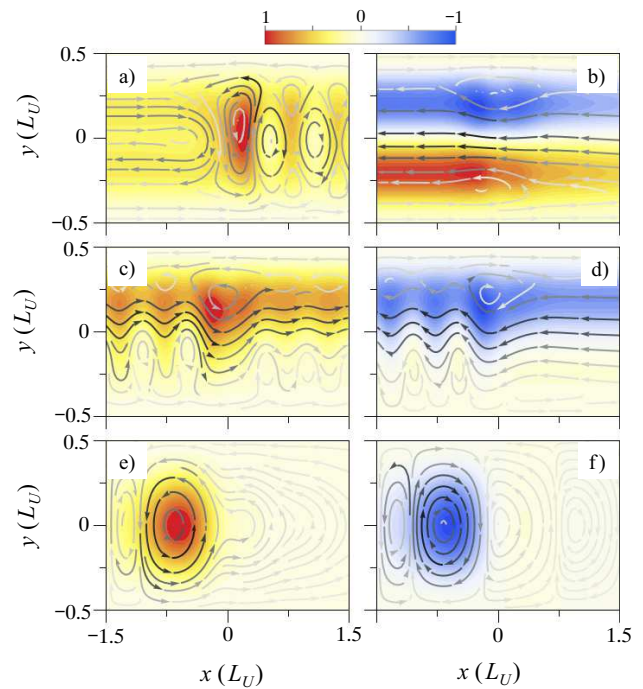}
}
\caption{Probability (color plot) and current (vector plot) for spatial distributions of 
quasiparticles (left) and charge (right). 
The results have been scaled dividing by the corresponding maximum absolute value of each 
field.
Each row of panels corresponds to a different configuration. Panels a) and b) are obtained with $\Delta_B=12\, E_U$ and $(\theta,\phi)=(80^{\rm o},10^{\rm o})$, when a Majorana mode is clearly present.
Panels c) are d) are for the same $\Delta_B$ but in a more tilted orientation $(\theta,\phi)=(80^{\rm o},15^{\rm o})$, when the Majorana mode is lost due to the projection rule. 
Panels e) and f) are with $(\theta,\phi)=(80^{\rm o},0)$ 
and a reduced field $\Delta_B=E_U$, when the Majorana is lost due to the critical field rule.
Rest of parameters: $\Delta_0=3\, E_U$, $\alpha=2\, E_U L_U$, $g^*=15$, $m^*=0.033\, m_e$, $V_b=5\; E_U$, $L_b=0.3\, L_U$, $\mu=0$ (a-d), $\mu=3E_U$ (e, f). 
}
\label{F1}
\end{figure}

\section{Results}

As in Ref.\ \onlinecite{Os15} we consider a unit system characterized by 
a length unit $L_U$ and a corresponding energy unit defined with $\hbar$ and $m$ as 
$E_U=\hbar^2/mL_U^2$.
Our natural choice is $L_U=L_y$, the transverse width of the 2D stripe.
With $L_U=150$ nm and $m=0.033m_e$, typical with InAs, it is
$E_U=  0.10$ meV. Below, we will also assume other typical values as
$\alpha= 2\, E_UL_U$, $\Delta_0=3\, E_U$. 
Also, from $\Delta_B = g^* \mu_B B/2$, with 
$\mu_B$ the Bohr magneton and $g^*=15$ (gyromagnetic factor),
we may obtain 
the magnetic field modulus from $\Delta_B$ as
$B=0.23 (\Delta_B/E_U)\, {\rm T}$. 

\subsection{Characteristic distributions}

Figure \ref{F1} presents typical results of quasiparticle/charge density/current 
distributions in a magnetic field with vertical component when a Majorana mode is 
present (panels a, b) and when it is destroyed by either  tilting the field with respect
to the $z$ axis (panels c, d) or by decreasing the field intensity (panels e, f).
A potential barrier of a moderate height is present at the junction. 
In Fig.\ \ref{F1}a the quasiparticle density 
clearly reflects  
the presence of the Majorana on the R (right, superconductor) side while the quasiparticle current
shows a U-turn shape and a sequence of vortices  on the L and R sides, respectively.
The corresponding charge distributions (Fig.\ \ref{F1}b) are markedly different.
The charge density is characterized by charge accumulations in the normal side of the stripe,
of reversed signs for opposite sides, that fade away when entering the superconducting side.
Being charge neutral, the Majorana
leaves only minor density distortions localized close the junction. The charge current
in Fig.\ \ref{F1}b is $y$-symmetric  in the L contact and it slowly vanishes 
when entering the R lead. This scenario is understood in terms of the Andreev reflection
of incident electrons into holes made possible even in presence of the junction barrier 
by the Majorana mode.

\begin{figure}[t]
\centering
\resizebox{0.325\textwidth}{!}{
	\includegraphics{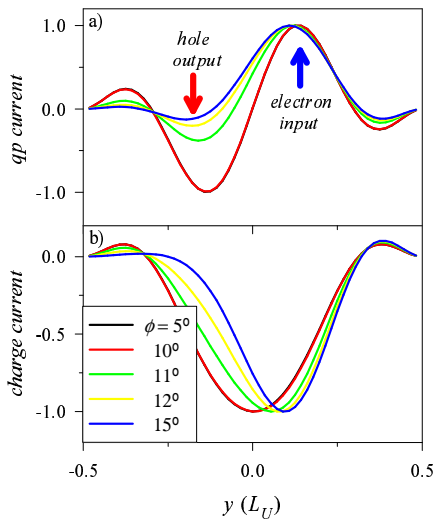}
}
\caption{
Transverse cuts of the quasiparticle (a) and charge (b) currents in the normal lead and its
evolution with azimuthal angle $\phi$. The same parameters of Fig.\ \ref{F1}c and \ref{F1}d have been used, except of $\phi$ that is varied as specified. The results are scaled by the corresponding maximum 
absolute values.}
\label{F2}
\end{figure}

Results in Figs.\ \ref{F1}c and \ref{F1}d show the evolution when further tilting the magnetic field in the lateral direction, keeping the vertical component constant. Increasing the azimuthal angle $\phi$ the projection rule mentioned in Sec.\ \ref{intro} sets up a limit for the stability of the Majorana state, when the energy gap closes on the 
superconducting side and propagating modes emerge towards the right.\cite{note}
Crossing this critical $\phi$, Andreev reflection is strongly quenched in front of normal transmission and reflection.
This change has a strong influence on the spatial distributions of Figs.\ \ref{F1}c and \ref{F1}d
with respect to the preceding situation with the Majorana mode.
Quasiparticle 
and charge distributions are now basically due to electronic states only, and hence, they are rather 
similar once the negative sign of the electron charge is taken into account. Normal reflection
causes wiggling of the reflected charge current (Fig.\ \ref{F1}d) but the most important effect
is the side shift of the current, in sharp contrast with the situation of
Fig.\ \ref{F1}b. This shift is caused by the vertical component of the field and, thus, it is 
qualitatively similar to the transmission by edge modes of the quantum Hall effect.

Figures \ref{F1}e and \ref{F1}f show the results when the field intensity is reduced below the critical value for the 
presence of a Majorana mode. 
A chemical potential $\mu=3E_U$ has also been used to allow for propagating states in the normal lead.
In this situation, normal transmission 
is greatly quenched in front of normal reflection by the barrier and the pattern of interferences on the normal side is enhanced. Indeed, a sequence of vortices in the quasiparticle and charge densities are seen in the L lead in Figs.\ \ref{F1}e  and \ref{F1}f.

More clear transverse cuts of quasiparticle and charge currents in the asymptotic region of the normal lead are shown in Fig.\ \ref{F2}. Panel \ref{F2}a (quasiparticle current) shows 
for $\phi=0$ electrons and holes flowing along opposite transverse sides and the disappearance 
of the holes when increasing $\phi$. Panel \ref{F2}b shows the above mentioned side shift of the charge current when increasing $\phi$. The profile is symmetric in presence of the Majorana, until
$\phi$ exceeds a critical value of $\approx 11$ degrees, at which point the 
$y$ inversion 
symmetry is gradually lost.

\begin{figure}[t]
\centering
\resizebox{0.5\textwidth}{!}{
	\includegraphics{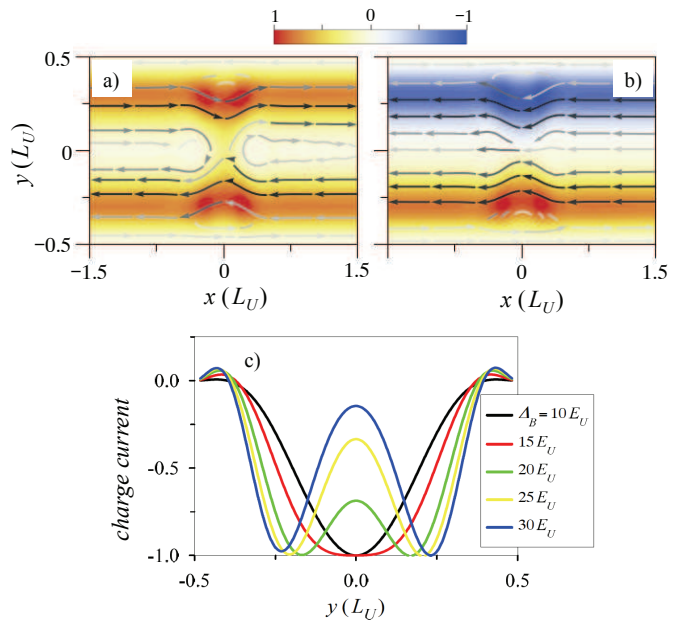}
}
\caption{
High magnetic field results, when edge modes are formed and the Majorana is split.
Similarly to Fig.\ \ref{F1},  panels a) and b) correspond to quasiparticle and charge spatial 
distributions for $\Delta_B=30\; E_U$ and $(\theta,\phi)=(82^{\rm o}, 0)$.
Panel c) shows the evolution with $\Delta_B$ of a transverse cut in the charge current deep in the normal 
lead.
}
\label{F3}
\end{figure}

With very large magnetic fields, when $l_z<L_y$ ($l_z$ and $L_y$ being  the  magnetic length and the stripe width, respectively) the modes become increasingly edge-like. In this situation, we find that the 
decay of the evanescent modes towards the R lead is extremely small, the density and current distributions getting attached to the sides of the stripe and penetrating the superconductor side (Fig.\ \ref{F3}). At such high magnetic fields the Majorana is not leaving a noticeable fingerprint on the spatial distributions, although the charge current profile is still $y$-symmetric but characterized by 
two peaks on the two lateral edges  (Fig.\ \ref{F3}c). The quasiparticle distribution, Fig.\ \ref{F3}a, is 
also showing how the Majorana peak splits into two because of the large magnetic tendency to attach quasiparticles along the side edges.

\begin{figure}[t]
\centering
\resizebox{0.35\textwidth}{!}{
	\includegraphics{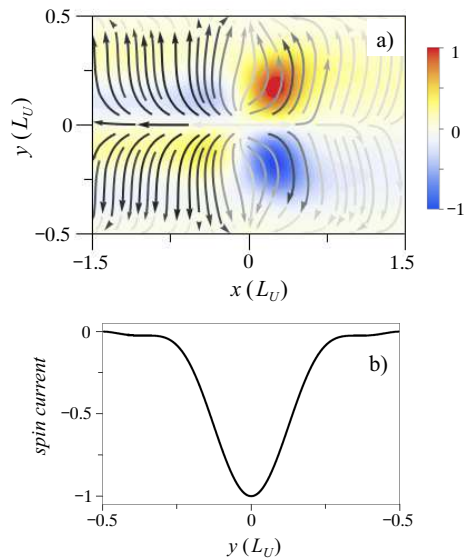}
}
\caption{
a) Distribution of spin density $\rho_{sp}$ (color plot) and spin current $\vec\jmath_{sp}$ (arrow plot).
b) transverse cut of $\vec{\jmath}_{sp}$ in the L lead. The same parameters of 
Fig.\ \ref{F1}a have been used.
The results are scaled by the corresponding maximum 
absolute values.
} 
\label{F4}
\end{figure}

\subsection{Spin distributions}

Spin magnetization distributions $\rho_{sp}(x,y)$
and $\vec{\jmath}_{sp}(x,y)$ are shown in Fig.\ \ref{F4} for a selected case. They correspond to the presence of the Majorana state shown in Fig.\ \ref{F1}a, for a field with $\theta=80^{\rm o}$ and $\phi=0$. Notice that accumulations of spins of different signs occur on the two
sides of the stripe. These accumulations are particularly strong on the position of the Majorana
state and are due to the spin flip between incident and reflected quasiparticles, induced by the junction.
Remarkably, such spin-flip  Andreev reflection is 
a manifestation of an effective spin triplet pairing on the $R$ lead, due to the combination of Zeeman, pairing and Rashba interactions. The junction is then acting as a source of spin current $\vec\jmath_{sp}$ towards the normal lead and whose transverse profile is $y$-symmetric (Fig.\ \ref{F4}b).  The spin current is obviously not conserved which is not surprising since, as already mentioned, Eq.\ (\ref{E18})
is not fulfilled with spin distributions.

\begin{figure}[t]
\centering
\resizebox{0.49\textwidth}{!}{
	\includegraphics{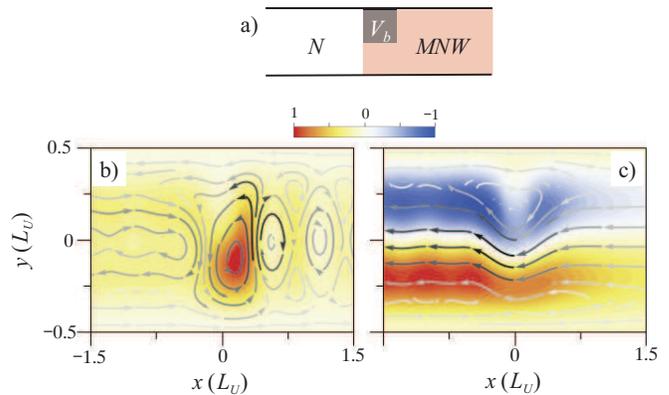}
}
\caption{Panel a) is a sketch of an asymmetric junction barrier. Panels b) and c) show the same as Fig.\ \ref{F1}a
and  \ref{F1}b but for an asymmetric barrier covering only half of the 
stripe width.
}
\label{F5}
\end{figure}

\subsection{Asymmetries}

We have explored the situation in which the junction barrier contains $y$-aymmetries (Fig.\ \ref{F5}). In this case the distributions $\rho_{qp}$ and $\vec\jmath_{qp}$, reflecting the position of the Majorana mode, are distorted
with respect to those of a symmetric barrier (Figs.\ \ref{F1}a and \ref{F1}b). The distortion is also seen in the charge distributions
$\rho_c$ and $\vec\jmath_{c}$
shown in Fig.\ \ref{F5}b. These modifications, however, are restricted  to the vicinity
of the junction, the distributions  rapidly recovering the same shape of the symmetric barrier when going towards the normal lead. In particular,  we find a  
$y$-symmetric charge current in the L lead in spite of the presence of an asymmetric barrier. The lack of dependence of the asymptotic distributions on specific details of the barrier agrees with the expected robust topological behavior of the junction.

\section{Conclusions}

We have calculated  density/current spatial distributions of
quasiparticle/charge/spin
in a 2D junction 
between a normal and a hybrid superconducting lead. The current distributions in the normal lead
present a strong variation with the topological phase of the superconducting lead. When a Majorana mode
is present the charge current is $y$-symmetric due to the dominant Andreev reflection processes.
Tilting the field laterally, while keeping the vertical component fixed, the current is suddenly displaced to the lead lateral side when crossing the Majorana phase boundary. Decreasing the field magnitude, 
Andreev reflection is replaced by normal reflection due to the potential barrier and the charge current presents  typical interference patterns.

The Majorana mode causes spin density accumulations of opposite signs as well as the emergence of 
a $y$-symmetric spin current propagating in the normal lead. The spin flip mechanism can be understood as a manifestation of an effective triplet pairing in the Majorana nanowire. Finally, the current distributions in the asymptotic parts of the normal lead are not affected by asymmetries of the junction 
barrier, as expected for a robust topological behaviour.  

\begin{acknowledgments}
This work was funded by MINECO-Spain, grant FIS2014-52564.
Discussions with C.\ Beenakker and R. S\'anchez are gratefully acknowledged.
J. O. acknowledges a PhD grant from the University of the Balearic Islands.
\end{acknowledgments}

\bibliographystyle{apsrev4-1}
\bibliography{Articulo_5L}

%merlin.mbs 2010-03-15 4.21a (PWD, AO, DPC)
\begin{thebibliography}{39}%
\makeatletter
\providecommand \@ifxundefined [1]{%
 \@ifx{#1\undefined}
}%
\providecommand \@ifnum [1]{%
 \ifnum #1\expandafter \@firstoftwo
 \else \expandafter \@secondoftwo
 \fi
}%
\providecommand \@ifx [1]{%
 \ifx #1\expandafter \@firstoftwo
 \else \expandafter \@secondoftwo
 \fi
}%
\providecommand \natexlab [1]{#1}%
\providecommand \enquote  [1]{``#1''}%
\providecommand \bibnamefont  [1]{#1}%
\providecommand \bibfnamefont [1]{#1}%
\providecommand \citenamefont [1]{#1}%
\providecommand \href@noop [0]{\@secondoftwo}%
\providecommand \href [0]{\begingroup \@sanitize@url \@href}%
\providecommand \@href[1]{\@@startlink{#1}\@@href}%
\providecommand \@@href[1]{\endgroup#1\@@endlink}%
\providecommand \@sanitize@url [0]{\catcode `\\12\catcode `\$12\catcode
  `\&12\catcode `\#12\catcode `\^12\catcode `\_12\catcode `\%12\relax}%
\providecommand \@@startlink[1]{}%
\providecommand \@@endlink[0]{}%
\providecommand \url  [0]{\begingroup\@sanitize@url \@url }%
\providecommand \@url [1]{\endgroup\@href {#1}{\urlprefix }}%
\providecommand \urlprefix  [0]{URL }%
\providecommand \Eprint [0]{\href }%
\@ifxundefined \urlstyle {%
  \providecommand \doi  [0]{\begingroup \@sanitize@url \@doi}%
  \providecommand \@doi [1]{\endgroup \@@startlink {\doibase
  #1}doi:\discretionary {}{}{}#1\@@endlink }%
}{%
  \providecommand \doi  [0]{doi:\discretionary{}{}{}\begingroup
  \urlstyle{rm}\Url }%
}%
\providecommand \doibase [0]{http://dx.doi.org/}%
\providecommand \Doi [0]{\begingroup \@sanitize@url \@Doi }%
\providecommand \@Doi  [1]{\endgroup\@@startlink{\doibase#1}\@@Doi}%
\providecommand \@@Doi [1]{#1\@@endlink}%
\providecommand \selectlanguage [0]{\@gobble}%
\providecommand \bibinfo  [0]{\@secondoftwo}%
\providecommand \bibfield  [0]{\@secondoftwo}%
\providecommand \translation [1]{[#1]}%
\providecommand \BibitemOpen [0]{}%
\providecommand \bibitemStop [0]{}%
\providecommand \bibitemNoStop [0]{.\EOS\space}%
\providecommand \EOS [0]{\spacefactor3000\relax}%
\providecommand \BibitemShut  [1]{\csname bibitem#1\endcsname}%
%</preamble>
\bibitem [{\citenamefont {Alicea}(2012)}]{Alicea}%
  \BibitemOpen
  \bibfield  {author} {\bibinfo {author} {\bibfnamefont {Jason}\ \bibnamefont
  {Alicea}},\ }\bibfield  {title} {\enquote {\bibinfo {title} {New directions
  in the pursuit of {Majorana} fermions in solid state systems},}\ }\href@noop
  {} {\bibfield  {journal} {\bibinfo  {journal} {Rep.\ Prog.\ Phys.},\ }\textbf
  {\bibinfo {volume} {75}},\ \bibinfo {pages} {076501} (\bibinfo {year}
  {2012})}\BibitemShut {NoStop}%
\bibitem [{\citenamefont {Beenakker}(2013)}]{Beenakker}%
  \BibitemOpen
  \bibfield  {author} {\bibinfo {author} {\bibfnamefont {C.~W.~J.}\
  \bibnamefont {Beenakker}},\ }\bibfield  {title} {\enquote {\bibinfo {title}
  {Search for {Majorana} fermions in superconductors},}\ }\Doi
  {10.1146/annurev-conmatphys-030212-184337} {\bibfield  {journal} {\bibinfo
  {journal} {Annual Review of Condensed Matter Physics},\ }\textbf {\bibinfo
  {volume} {4}},\ \bibinfo {pages} {113--136} (\bibinfo {year}
  {2013})}\BibitemShut {NoStop}%
\bibitem [{\citenamefont {Stanescu}\ and\ \citenamefont
  {Tewari}(2013)}]{StanescuREV}%
  \BibitemOpen
  \bibfield  {author} {\bibinfo {author} {\bibfnamefont {T.~D.}\ \bibnamefont
  {Stanescu}}\ and\ \bibinfo {author} {\bibfnamefont {S.}~\bibnamefont
  {Tewari}},\ }\bibfield  {title} {\enquote {\bibinfo {title} {Majorana
  fermions in semiconductor nanowires: fundamentals, modeling, and
  experiment},}\ }\href {http://stacks.iop.org/0953-8984/25/i=23/a=233201}
  {\bibfield  {journal} {\bibinfo  {journal} {Journal of Physics: Condensed
  Matter},\ }\textbf {\bibinfo {volume} {25}},\ \bibinfo {pages} {233201}
  (\bibinfo {year} {2013})}\BibitemShut {NoStop}%
\bibitem [{\citenamefont {Franz}(2013)}]{Franz}%
  \BibitemOpen
  \bibfield  {author} {\bibinfo {author} {\bibfnamefont {Marcel}\ \bibnamefont
  {Franz}},\ }\bibfield  {title} {\enquote {\bibinfo {title} {Majorana's
  wires},}\ }\href {http://dx.doi.org/10.1038/nnano.2013.33} {\bibfield
  {journal} {\bibinfo  {journal} {Nat Nano},\ }\textbf {\bibinfo {volume}
  {8}},\ \bibinfo {pages} {149--152} (\bibinfo {year} {2013})}\BibitemShut
  {NoStop}%
\bibitem [{\citenamefont {Fu}\ and\ \citenamefont {Kane}(2008)}]{Fu}%
  \BibitemOpen
  \bibfield  {author} {\bibinfo {author} {\bibfnamefont {Liang}\ \bibnamefont
  {Fu}}\ and\ \bibinfo {author} {\bibfnamefont {C.~L.}\ \bibnamefont {Kane}},\
  }\bibfield  {title} {\enquote {\bibinfo {title} {Superconducting proximity
  effect and {Majorana} fermions at the surface of a topological insulator},}\
  }\Doi {10.1103/PhysRevLett.100.096407} {\bibfield  {journal} {\bibinfo
  {journal} {Phys. Rev. Lett.},\ }\textbf {\bibinfo {volume} {100}},\ \bibinfo
  {pages} {096407} (\bibinfo {year} {2008})}\BibitemShut {NoStop}%
\bibitem [{\citenamefont {Lutchyn}\ \emph {et~al.}(2010)\citenamefont
  {Lutchyn}, \citenamefont {Sau},\ and\ \citenamefont {Das~Sarma}}]{Lutchyn}%
  \BibitemOpen
  \bibfield  {author} {\bibinfo {author} {\bibfnamefont {Roman~M.}\
  \bibnamefont {Lutchyn}}, \bibinfo {author} {\bibfnamefont {Jay~D.}\
  \bibnamefont {Sau}}, \ and\ \bibinfo {author} {\bibfnamefont
  {S.}~\bibnamefont {Das~Sarma}},\ }\bibfield  {title} {\enquote {\bibinfo
  {title} {Majorana fermions and a topological phase transition in
  semiconductor-superconductor heterostructures},}\ }\Doi
  {10.1103/PhysRevLett.105.077001} {\bibfield  {journal} {\bibinfo  {journal}
  {Phys. Rev. Lett.},\ }\textbf {\bibinfo {volume} {105}},\ \bibinfo {pages}
  {077001} (\bibinfo {year} {2010})}\BibitemShut {NoStop}%
\bibitem [{\citenamefont {Oreg}\ \emph {et~al.}(2010)\citenamefont {Oreg},
  \citenamefont {Refael},\ and\ \citenamefont {von Oppen}}]{Oreg}%
  \BibitemOpen
  \bibfield  {author} {\bibinfo {author} {\bibfnamefont {Yuval}\ \bibnamefont
  {Oreg}}, \bibinfo {author} {\bibfnamefont {Gil}\ \bibnamefont {Refael}}, \
  and\ \bibinfo {author} {\bibfnamefont {Felix}\ \bibnamefont {von Oppen}},\
  }\bibfield  {title} {\enquote {\bibinfo {title} {Helical liquids and
  {Majorana} bound states in quantum wires},}\ }\Doi
  {10.1103/PhysRevLett.105.177002} {\bibfield  {journal} {\bibinfo  {journal}
  {Phys. Rev. Lett.},\ }\textbf {\bibinfo {volume} {105}},\ \bibinfo {pages}
  {177002} (\bibinfo {year} {2010})}\BibitemShut {NoStop}%
\bibitem [{\citenamefont {Mourik}\ \emph {et~al.}(2012)\citenamefont {Mourik},
  \citenamefont {Zuo}, \citenamefont {Frolov}, \citenamefont {Plissard},
  \citenamefont {Bakkers},\ and\ \citenamefont {Kouwenhoven}}]{Mourik}%
  \BibitemOpen
  \bibfield  {author} {\bibinfo {author} {\bibfnamefont {V.}~\bibnamefont
  {Mourik}}, \bibinfo {author} {\bibfnamefont {K.}~\bibnamefont {Zuo}},
  \bibinfo {author} {\bibfnamefont {S.M.}\ \bibnamefont {Frolov}}, \bibinfo
  {author} {\bibfnamefont {S.R.}\ \bibnamefont {Plissard}}, \bibinfo {author}
  {\bibfnamefont {E.P.A.M.}\ \bibnamefont {Bakkers}}, \ and\ \bibinfo {author}
  {\bibfnamefont {L.P.}\ \bibnamefont {Kouwenhoven}},\ }\bibfield  {title}
  {\enquote {\bibinfo {title} {Signatures of {Majorana} fermions in hybrid
  superconductor-semiconductor nanowire devices},}\ }\href@noop {} {\bibfield
  {journal} {\bibinfo  {journal} {Science},\ }\textbf {\bibinfo {volume}
  {336}},\ \bibinfo {pages} {1003--1007} (\bibinfo {year} {2012})}\BibitemShut
  {NoStop}%
\bibitem [{\citenamefont {Deng}\ \emph {et~al.}(2012)\citenamefont {Deng},
  \citenamefont {Yu}, \citenamefont {Huang}, \citenamefont {Larsson},
  \citenamefont {Caroff},\ and\ \citenamefont {Xu}}]{Deng}%
  \BibitemOpen
  \bibfield  {author} {\bibinfo {author} {\bibfnamefont {M.~T.}\ \bibnamefont
  {Deng}}, \bibinfo {author} {\bibfnamefont {C.~L.}\ \bibnamefont {Yu}},
  \bibinfo {author} {\bibfnamefont {G.~Y.}\ \bibnamefont {Huang}}, \bibinfo
  {author} {\bibfnamefont {M.}~\bibnamefont {Larsson}}, \bibinfo {author}
  {\bibfnamefont {P.}~\bibnamefont {Caroff}}, \ and\ \bibinfo {author}
  {\bibfnamefont {H.~Q.}\ \bibnamefont {Xu}},\ }\bibfield  {title} {\enquote
  {\bibinfo {title} {Anomalous zero-bias conductance peak in a {Nb-InSb}
  nanowire {Nb} hybrid device},}\ }\Doi {10.1021/nl303758w} {\bibfield
  {journal} {\bibinfo  {journal} {Nano Letters},\ }\textbf {\bibinfo {volume}
  {12}},\ \bibinfo {pages} {6414--6419} (\bibinfo {year} {2012})},\ \bibinfo
  {note} {pMID: 23181691},\ \Eprint
  {http://arxiv.org/abs/http://dx.doi.org/10.1021/nl303758w}
  {http://dx.doi.org/10.1021/nl303758w} \BibitemShut {NoStop}%
\bibitem [{\citenamefont {Das}\ \emph {et~al.}(2012)\citenamefont {Das},
  \citenamefont {Ronen}, \citenamefont {Most}, \citenamefont {Oreg},
  \citenamefont {Heiblum},\ and\ \citenamefont {Shtrikman}}]{Das}%
  \BibitemOpen
  \bibfield  {author} {\bibinfo {author} {\bibfnamefont {Anindya}\ \bibnamefont
  {Das}}, \bibinfo {author} {\bibfnamefont {Yuval}\ \bibnamefont {Ronen}},
  \bibinfo {author} {\bibfnamefont {Yonatan}\ \bibnamefont {Most}}, \bibinfo
  {author} {\bibfnamefont {Yuval}\ \bibnamefont {Oreg}}, \bibinfo {author}
  {\bibfnamefont {Moty}\ \bibnamefont {Heiblum}}, \ and\ \bibinfo {author}
  {\bibfnamefont {Hadas}\ \bibnamefont {Shtrikman}},\ }\bibfield  {title}
  {\enquote {\bibinfo {title} {Zero-bias peaks and splitting in an {Al-InAs}
  nanowire topological superconductor as a signature of {Majorana}
  {Fermions}},}\ }\href {http://dx.doi.org/10.1038/nphys2479} {\bibfield
  {journal} {\bibinfo  {journal} {Nat Phys},\ }\textbf {\bibinfo {volume}
  {8}},\ \bibinfo {pages} {887--895} (\bibinfo {year} {2012})}\BibitemShut
  {NoStop}%
\bibitem [{\citenamefont {Finck}\ \emph {et~al.}(2013)\citenamefont {Finck},
  \citenamefont {Van~Harlingen}, \citenamefont {Mohseni}, \citenamefont
  {Jung},\ and\ \citenamefont {Li}}]{Finck}%
  \BibitemOpen
  \bibfield  {author} {\bibinfo {author} {\bibfnamefont {A.~D.~K.}\
  \bibnamefont {Finck}}, \bibinfo {author} {\bibfnamefont {D.~J.}\ \bibnamefont
  {Van~Harlingen}}, \bibinfo {author} {\bibfnamefont {P.~K.}\ \bibnamefont
  {Mohseni}}, \bibinfo {author} {\bibfnamefont {K.}~\bibnamefont {Jung}}, \
  and\ \bibinfo {author} {\bibfnamefont {X.}~\bibnamefont {Li}},\ }\bibfield
  {title} {\enquote {\bibinfo {title} {Anomalous modulation of a zero-bias peak
  in a hybrid nanowire-superconductor device},}\ }\Doi
  {10.1103/PhysRevLett.110.126406} {\bibfield  {journal} {\bibinfo  {journal}
  {Phys. Rev. Lett.},\ }\textbf {\bibinfo {volume} {110}},\ \bibinfo {pages}
  {126406} (\bibinfo {year} {2013})}\BibitemShut {NoStop}%
\bibitem [{\citenamefont {Zhang}\ \emph {et~al.}(2016)\citenamefont {Zhang},
  \citenamefont {G\"ul}, \citenamefont {Conesa-Boj}, \citenamefont {Zuo},
  \citenamefont {Mourik}, \citenamefont {de~Vries}, \citenamefont {van Veen},
  \citenamefont {van Woerkom}, \citenamefont {Nowak}, \citenamefont {Wimmer},
  \citenamefont {Car}, \citenamefont {Plissard}, \citenamefont {Bakkers},
  \citenamefont {Quintero-P\'erez}, \citenamefont {Goswami}, \citenamefont
  {Watanabe}, \citenamefont {Taniguchi},\ and\ \citenamefont
  {Kouwenhoven}}]{Zhang}%
  \BibitemOpen
  \bibfield  {author} {\bibinfo {author} {\bibfnamefont {Hao}\ \bibnamefont
  {Zhang}}, \bibinfo {author} {\bibfnamefont {\"Onder}\ \bibnamefont {G\"ul}},
  \bibinfo {author} {\bibfnamefont {Sonia}\ \bibnamefont {Conesa-Boj}},
  \bibinfo {author} {\bibfnamefont {Kun}\ \bibnamefont {Zuo}}, \bibinfo
  {author} {\bibfnamefont {Vincent}\ \bibnamefont {Mourik}}, \bibinfo {author}
  {\bibfnamefont {Folkert~K.}\ \bibnamefont {de~Vries}}, \bibinfo {author}
  {\bibfnamefont {Jasper}\ \bibnamefont {van Veen}}, \bibinfo {author}
  {\bibfnamefont {David~J.}\ \bibnamefont {van Woerkom}}, \bibinfo {author}
  {\bibfnamefont {Micha~P.}\ \bibnamefont {Nowak}}, \bibinfo {author}
  {\bibfnamefont {Michael}\ \bibnamefont {Wimmer}}, \bibinfo {author}
  {\bibfnamefont {Diana}\ \bibnamefont {Car}}, \bibinfo {author} {\bibfnamefont
  {S\'ebastien}\ \bibnamefont {Plissard}}, \bibinfo {author} {\bibfnamefont
  {Erik P. A.~M.}\ \bibnamefont {Bakkers}}, \bibinfo {author} {\bibfnamefont
  {Marina}\ \bibnamefont {Quintero-P\'erez}}, \bibinfo {author} {\bibfnamefont
  {Srijit}\ \bibnamefont {Goswami}}, \bibinfo {author} {\bibfnamefont {Kenji}\
  \bibnamefont {Watanabe}}, \bibinfo {author} {\bibfnamefont {Takashi}\
  \bibnamefont {Taniguchi}}, \ and\ \bibinfo {author} {\bibfnamefont {Leo~P.}\
  \bibnamefont {Kouwenhoven}},\ }\bibfield  {title} {\enquote {\bibinfo {title}
  {Ballistic {Majorana} nanowire devices},}\ }\href@noop {} {\bibfield
  {journal} {\bibinfo  {journal} {arXiv:1603.04069}} (\bibinfo {year}
  {2016})}\BibitemShut {NoStop}%
\bibitem [{\citenamefont {Lutchyn}\ \emph {et~al.}(2011)\citenamefont
  {Lutchyn}, \citenamefont {Stanescu},\ and\ \citenamefont
  {Das~Sarma}}]{Lutchyn2}%
  \BibitemOpen
  \bibfield  {author} {\bibinfo {author} {\bibfnamefont {Roman~M.}\
  \bibnamefont {Lutchyn}}, \bibinfo {author} {\bibfnamefont {Tudor~D.}\
  \bibnamefont {Stanescu}}, \ and\ \bibinfo {author} {\bibfnamefont
  {S.}~\bibnamefont {Das~Sarma}},\ }\bibfield  {title} {\enquote {\bibinfo
  {title} {Search for {Majorana Fermions} in multiband semiconducting
  nanowires},}\ }\Doi {10.1103/PhysRevLett.106.127001} {\bibfield  {journal}
  {\bibinfo  {journal} {Phys. Rev. Lett.},\ }\textbf {\bibinfo {volume}
  {106}},\ \bibinfo {pages} {127001} (\bibinfo {year} {2011})}\BibitemShut
  {NoStop}%
\bibitem [{\citenamefont {Lutchyn}\ and\ \citenamefont
  {Fisher}(2011)}]{Lutchyn3}%
  \BibitemOpen
  \bibfield  {author} {\bibinfo {author} {\bibfnamefont {Roman~M.}\
  \bibnamefont {Lutchyn}}\ and\ \bibinfo {author} {\bibfnamefont {Matthew
  P.~A.}\ \bibnamefont {Fisher}},\ }\bibfield  {title} {\enquote {\bibinfo
  {title} {Interacting topological phases in multiband nanowires},}\ }\Doi
  {10.1103/PhysRevB.84.214528} {\bibfield  {journal} {\bibinfo  {journal}
  {Phys. Rev. B},\ }\textbf {\bibinfo {volume} {84}},\ \bibinfo {pages}
  {214528} (\bibinfo {year} {2011})}\BibitemShut {NoStop}%
\bibitem [{\citenamefont {Stanescu}\ \emph {et~al.}(2011)\citenamefont
  {Stanescu}, \citenamefont {Lutchyn},\ and\ \citenamefont
  {Das~Sarma}}]{Stanescu}%
  \BibitemOpen
  \bibfield  {author} {\bibinfo {author} {\bibfnamefont {Tudor~D.}\
  \bibnamefont {Stanescu}}, \bibinfo {author} {\bibfnamefont {Roman~M.}\
  \bibnamefont {Lutchyn}}, \ and\ \bibinfo {author} {\bibfnamefont
  {S.}~\bibnamefont {Das~Sarma}},\ }\bibfield  {title} {\enquote {\bibinfo
  {title} {Majorana fermions in semiconductor nanowires},}\ }\Doi
  {10.1103/PhysRevB.84.144522} {\bibfield  {journal} {\bibinfo  {journal}
  {Phys. Rev. B},\ }\textbf {\bibinfo {volume} {84}},\ \bibinfo {pages}
  {144522} (\bibinfo {year} {2011})}\BibitemShut {NoStop}%
\bibitem [{\citenamefont {Potter}\ and\ \citenamefont {Lee}(2010)}]{Potter}%
  \BibitemOpen
  \bibfield  {author} {\bibinfo {author} {\bibfnamefont {Andrew~C.}\
  \bibnamefont {Potter}}\ and\ \bibinfo {author} {\bibfnamefont {Patrick~A.}\
  \bibnamefont {Lee}},\ }\bibfield  {title} {\enquote {\bibinfo {title}
  {Multichannel generalization of {Kitaev's} {Majorana} end states and a
  practical route to realize them in thin films},}\ }\Doi
  {10.1103/PhysRevLett.105.227003} {\bibfield  {journal} {\bibinfo  {journal}
  {Phys. Rev. Lett.},\ }\textbf {\bibinfo {volume} {105}},\ \bibinfo {pages}
  {227003} (\bibinfo {year} {2010})}\BibitemShut {NoStop}%
\bibitem [{\citenamefont {Potter}\ and\ \citenamefont {Lee}(2011)}]{Potter2}%
  \BibitemOpen
  \bibfield  {author} {\bibinfo {author} {\bibfnamefont {Andrew~C.}\
  \bibnamefont {Potter}}\ and\ \bibinfo {author} {\bibfnamefont {Patrick~A.}\
  \bibnamefont {Lee}},\ }\bibfield  {title} {\enquote {\bibinfo {title}
  {Majorana end states in multiband microstructures with {Rashba} spin-orbit
  coupling},}\ }\Doi {10.1103/PhysRevB.83.094525} {\bibfield  {journal}
  {\bibinfo  {journal} {Phys. Rev. B},\ }\textbf {\bibinfo {volume} {83}},\
  \bibinfo {pages} {094525} (\bibinfo {year} {2011})}\BibitemShut {NoStop}%
\bibitem [{\citenamefont {Law}\ and\ \citenamefont {Lee}(2011)}]{Law11}%
  \BibitemOpen
  \bibfield  {author} {\bibinfo {author} {\bibfnamefont {K.~T.}\ \bibnamefont
  {Law}}\ and\ \bibinfo {author} {\bibfnamefont {Patrick~A.}\ \bibnamefont
  {Lee}},\ }\bibfield  {title} {\enquote {\bibinfo {title} {Robustness of
  {Majorana} fermion induced fractional {Josephson} effect in multichannel
  superconducting wires},}\ }\Doi {10.1103/PhysRevB.84.081304} {\bibfield
  {journal} {\bibinfo  {journal} {Phys. Rev. B},\ }\textbf {\bibinfo {volume}
  {84}},\ \bibinfo {pages} {081304} (\bibinfo {year} {2011})}\BibitemShut
  {NoStop}%
\bibitem [{\citenamefont {Tewari}\ and\ \citenamefont {Sau}(2012)}]{Tew12}%
  \BibitemOpen
  \bibfield  {author} {\bibinfo {author} {\bibfnamefont {Sumanta}\ \bibnamefont
  {Tewari}}\ and\ \bibinfo {author} {\bibfnamefont {Jay~D.}\ \bibnamefont
  {Sau}},\ }\bibfield  {title} {\enquote {\bibinfo {title} {Topological
  invariants for spin-orbit coupled superconductor nanowires},}\ }\Doi
  {10.1103/PhysRevLett.109.150408} {\bibfield  {journal} {\bibinfo  {journal}
  {Phys. Rev. Lett.},\ }\textbf {\bibinfo {volume} {109}},\ \bibinfo {pages}
  {150408} (\bibinfo {year} {2012})}\BibitemShut {NoStop}%
\bibitem [{\citenamefont {San-Jose}\ \emph {et~al.}(2014)\citenamefont
  {San-Jose}, \citenamefont {Prada},\ and\ \citenamefont {Aguado}}]{sanjose}%
  \BibitemOpen
  \bibfield  {author} {\bibinfo {author} {\bibfnamefont {Pablo}\ \bibnamefont
  {San-Jose}}, \bibinfo {author} {\bibfnamefont {Elsa}\ \bibnamefont {Prada}},
  \ and\ \bibinfo {author} {\bibfnamefont {Ram\'on}\ \bibnamefont {Aguado}},\
  }\bibfield  {title} {\enquote {\bibinfo {title} {Mapping the topological
  phase diagram of multiband semiconductors with supercurrents},}\ }\Doi
  {10.1103/PhysRevLett.112.137001} {\bibfield  {journal} {\bibinfo  {journal}
  {Phys. Rev. Lett.},\ }\textbf {\bibinfo {volume} {112}},\ \bibinfo {pages}
  {137001} (\bibinfo {year} {2014})}\BibitemShut {NoStop}%
\bibitem [{\citenamefont {Wang}\ \emph {et~al.}(2015)\citenamefont {Wang},
  \citenamefont {Liu}, \citenamefont {Sun},\ and\ \citenamefont {Xie}}]{Wan15}%
  \BibitemOpen
  \bibfield  {author} {\bibinfo {author} {\bibfnamefont {Pei}\ \bibnamefont
  {Wang}}, \bibinfo {author} {\bibfnamefont {Jie}\ \bibnamefont {Liu}},
  \bibinfo {author} {\bibfnamefont {Qing-feng}\ \bibnamefont {Sun}}, \ and\
  \bibinfo {author} {\bibfnamefont {X.~C.}\ \bibnamefont {Xie}},\ }\bibfield
  {title} {\enquote {\bibinfo {title} {Identifying the topological
  superconducting phase in a multiband quantum wire},}\ }\Doi
  {10.1103/PhysRevB.91.224512} {\bibfield  {journal} {\bibinfo  {journal}
  {Phys. Rev. B},\ }\textbf {\bibinfo {volume} {91}},\ \bibinfo {pages}
  {224512} (\bibinfo {year} {2015})}\BibitemShut {NoStop}%
\bibitem [{\citenamefont {Wimmer}\ \emph {et~al.}(2011)\citenamefont {Wimmer},
  \citenamefont {Akhmerov}, \citenamefont {Dahlhaus},\ and\ \citenamefont
  {Beenakker}}]{Wim11}%
  \BibitemOpen
  \bibfield  {author} {\bibinfo {author} {\bibfnamefont {M}~\bibnamefont
  {Wimmer}}, \bibinfo {author} {\bibfnamefont {A~R}\ \bibnamefont {Akhmerov}},
  \bibinfo {author} {\bibfnamefont {J~P}\ \bibnamefont {Dahlhaus}}, \ and\
  \bibinfo {author} {\bibfnamefont {C~W~J}\ \bibnamefont {Beenakker}},\
  }\bibfield  {title} {\enquote {\bibinfo {title} {Quantum point contact as a
  probe of a topological superconductor},}\ }\href
  {http://stacks.iop.org/1367-2630/13/i=5/a=053016} {\bibfield  {journal}
  {\bibinfo  {journal} {New Journal of Physics},\ }\textbf {\bibinfo {volume}
  {13}},\ \bibinfo {pages} {053016} (\bibinfo {year} {2011})}\BibitemShut
  {NoStop}%
\bibitem [{\citenamefont {Lim}\ \emph {et~al.}(2013)\citenamefont {Lim},
  \citenamefont {Lopez},\ and\ \citenamefont {Serra}}]{Lim3}%
  \BibitemOpen
  \bibfield  {author} {\bibinfo {author} {\bibfnamefont {Jong~Soo}\
  \bibnamefont {Lim}}, \bibinfo {author} {\bibfnamefont {Rosa}\ \bibnamefont
  {Lopez}}, \ and\ \bibinfo {author} {\bibfnamefont {Lloren\c{c}}\ \bibnamefont
  {Serra}},\ }\bibfield  {title} {\enquote {\bibinfo {title} {Emergence of
  {Majorana} modes in cylindrical nanowires},}\ }\href
  {http://stacks.iop.org/0295-5075/103/i=3/a=37004} {\bibfield  {journal}
  {\bibinfo  {journal} {EPL (Europhysics Letters)},\ }\textbf {\bibinfo
  {volume} {103}},\ \bibinfo {pages} {37004} (\bibinfo {year}
  {2013})}\BibitemShut {NoStop}%
\bibitem [{\citenamefont {Osca}\ \emph {et~al.}(2014)\citenamefont {Osca},
  \citenamefont {L{\'o}pez},\ and\ \citenamefont {Serra}}]{Os14}%
  \BibitemOpen
  \bibfield  {author} {\bibinfo {author} {\bibfnamefont {Javier}\ \bibnamefont
  {Osca}}, \bibinfo {author} {\bibfnamefont {Rosa}\ \bibnamefont {L{\'o}pez}},
  \ and\ \bibinfo {author} {\bibfnamefont {Lloren{\c{c}}}\ \bibnamefont
  {Serra}},\ }\bibfield  {title} {\enquote {\bibinfo {title} {Majorana mode
  stacking, robustness and size effect in cylindrical nanowires},}\ }\Doi
  {10.1140/epjb/e2014-41091-8} {\bibfield  {journal} {\bibinfo  {journal} {The
  European Physical Journal B},\ }\textbf {\bibinfo {volume} {87}},\ \bibinfo
  {pages} {1--7} (\bibinfo {year} {2014})},\ ISSN \bibinfo {issn}
  {1434-6036}\BibitemShut {NoStop}%
\bibitem [{\citenamefont {Osca}\ and\ \citenamefont
  {Serra}(2015){\natexlab{a}}}]{Os15}%
  \BibitemOpen
  \bibfield  {author} {\bibinfo {author} {\bibfnamefont {Javier}\ \bibnamefont
  {Osca}}\ and\ \bibinfo {author} {\bibfnamefont {Lloren\c{c}}\ \bibnamefont
  {Serra}},\ }\bibfield  {title} {\enquote {\bibinfo {title} {Majorana states
  and magnetic orbital motion in planar hybrid nanowires},}\ }\Doi
  {10.1103/PhysRevB.91.235417} {\bibfield  {journal} {\bibinfo  {journal}
  {Phys. Rev. B},\ }\textbf {\bibinfo {volume} {91}},\ \bibinfo {pages}
  {235417} (\bibinfo {year} {2015}{\natexlab{a}})}\BibitemShut {NoStop}%
\bibitem [{\citenamefont {Nijholt}\ and\ \citenamefont
  {Akhmerov}(2016)}]{Nij16}%
  \BibitemOpen
  \bibfield  {author} {\bibinfo {author} {\bibfnamefont {Bas}\ \bibnamefont
  {Nijholt}}\ and\ \bibinfo {author} {\bibfnamefont {Anton~R.}\ \bibnamefont
  {Akhmerov}},\ }\bibfield  {title} {\enquote {\bibinfo {title} {Orbital effect
  of magnetic field on the {Majorana} phase diagram},}\ }\Doi
  {10.1103/PhysRevB.93.235434} {\bibfield  {journal} {\bibinfo  {journal}
  {Phys. Rev. B},\ }\textbf {\bibinfo {volume} {93}},\ \bibinfo {pages}
  {235434} (\bibinfo {year} {2016})}\BibitemShut {NoStop}%
\bibitem [{\citenamefont {Sedlmayr}\ \emph {et~al.}(2016)\citenamefont
  {Sedlmayr}, \citenamefont {Aguiar-Hualde},\ and\ \citenamefont
  {Bena}}]{Sed16}%
  \BibitemOpen
  \bibfield  {author} {\bibinfo {author} {\bibfnamefont {N.}~\bibnamefont
  {Sedlmayr}}, \bibinfo {author} {\bibfnamefont {J.~M.}\ \bibnamefont
  {Aguiar-Hualde}}, \ and\ \bibinfo {author} {\bibfnamefont {C.}~\bibnamefont
  {Bena}},\ }\bibfield  {title} {\enquote {\bibinfo {title} {Majorana bound
  states in open quasi-one-dimensional and two-dimensional systems with
  transverse {Rashba} coupling},}\ }\Doi {10.1103/PhysRevB.93.155425}
  {\bibfield  {journal} {\bibinfo  {journal} {Phys. Rev. B},\ }\textbf
  {\bibinfo {volume} {93}},\ \bibinfo {pages} {155425} (\bibinfo {year}
  {2016})}\BibitemShut {NoStop}%
\bibitem [{\citenamefont {Shabani}\ \emph {et~al.}(2016)\citenamefont
  {Shabani}, \citenamefont {Kjaergaard}, \citenamefont {Suominen},
  \citenamefont {Kim}, \citenamefont {Nichele}, \citenamefont {Pakrouski},
  \citenamefont {Stankevic}, \citenamefont {Lutchyn}, \citenamefont
  {Krogstrup}, \citenamefont {Feidenhans'l}, \citenamefont {Kraemer},
  \citenamefont {Nayak}, \citenamefont {Troyer}, \citenamefont {Marcus},\ and\
  \citenamefont {Palmstr\o{}m}}]{Sha15}%
  \BibitemOpen
  \bibfield  {author} {\bibinfo {author} {\bibfnamefont {J.}~\bibnamefont
  {Shabani}}, \bibinfo {author} {\bibfnamefont {M.}~\bibnamefont {Kjaergaard}},
  \bibinfo {author} {\bibfnamefont {H.~J.}\ \bibnamefont {Suominen}}, \bibinfo
  {author} {\bibfnamefont {Younghyun}\ \bibnamefont {Kim}}, \bibinfo {author}
  {\bibfnamefont {F.}~\bibnamefont {Nichele}}, \bibinfo {author} {\bibfnamefont
  {K.}~\bibnamefont {Pakrouski}}, \bibinfo {author} {\bibfnamefont
  {T.}~\bibnamefont {Stankevic}}, \bibinfo {author} {\bibfnamefont {R.~M.}\
  \bibnamefont {Lutchyn}}, \bibinfo {author} {\bibfnamefont {P.}~\bibnamefont
  {Krogstrup}}, \bibinfo {author} {\bibfnamefont {R.}~\bibnamefont
  {Feidenhans'l}}, \bibinfo {author} {\bibfnamefont {S.}~\bibnamefont
  {Kraemer}}, \bibinfo {author} {\bibfnamefont {C.}~\bibnamefont {Nayak}},
  \bibinfo {author} {\bibfnamefont {M.}~\bibnamefont {Troyer}}, \bibinfo
  {author} {\bibfnamefont {C.~M.}\ \bibnamefont {Marcus}}, \ and\ \bibinfo
  {author} {\bibfnamefont {C.~J.}\ \bibnamefont {Palmstr\o{}m}},\ }\bibfield
  {title} {\enquote {\bibinfo {title} {Two-dimensional epitaxial
  superconductor-semiconductor heterostructures: A platform for topological
  superconducting networks},}\ }\Doi {10.1103/PhysRevB.93.155402} {\bibfield
  {journal} {\bibinfo  {journal} {Phys. Rev. B},\ }\textbf {\bibinfo {volume}
  {93}},\ \bibinfo {pages} {155402} (\bibinfo {year} {2016})}\BibitemShut
  {NoStop}%
\bibitem [{\citenamefont {Kjaergaard}\ \emph {et~al.}(2016)\citenamefont
  {Kjaergaard}, \citenamefont {Nichele}, \citenamefont {Suominen},
  \citenamefont {Nowak}, \citenamefont {Wimmer}, \citenamefont {Akhmerov},
  \citenamefont {Folk}, \citenamefont {Flensberg}, \citenamefont {Shabani},
  \citenamefont {Palmstr¿m},\ and\ \citenamefont {Marcus}}]{Kja16}%
  \BibitemOpen
  \bibfield  {author} {\bibinfo {author} {\bibfnamefont {M.}~\bibnamefont
  {Kjaergaard}}, \bibinfo {author} {\bibfnamefont {F.}~\bibnamefont {Nichele}},
  \bibinfo {author} {\bibfnamefont {H.~J.}\ \bibnamefont {Suominen}}, \bibinfo
  {author} {\bibfnamefont {M.~P.}\ \bibnamefont {Nowak}}, \bibinfo {author}
  {\bibfnamefont {M.}~\bibnamefont {Wimmer}}, \bibinfo {author} {\bibfnamefont
  {A.~R.}\ \bibnamefont {Akhmerov}}, \bibinfo {author} {\bibfnamefont {J.~A.}\
  \bibnamefont {Folk}}, \bibinfo {author} {\bibfnamefont {K.}~\bibnamefont
  {Flensberg}}, \bibinfo {author} {\bibfnamefont {J.}~\bibnamefont {Shabani}},
  \bibinfo {author} {\bibfnamefont {C.~J.}\ \bibnamefont {Palmstr¿m}}, \ and\
  \bibinfo {author} {\bibfnamefont {C.~M.}\ \bibnamefont {Marcus}},\ }\bibfield
   {title} {\enquote {\bibinfo {title} {Quantized conductance doubling and hard
  gap in a two-dimensional semiconductor-superconductor heterostructure},}\
  }\href@noop {} {\bibfield  {journal} {\bibinfo  {journal}
  {arXiv:1603.01852v1}} (\bibinfo {year} {2016})}\BibitemShut {NoStop}%
\bibitem [{\citenamefont {Beenakker}(1992)}]{Bee92}%
  \BibitemOpen
  \bibfield  {author} {\bibinfo {author} {\bibfnamefont {C.~W.~J.}\
  \bibnamefont {Beenakker}},\ }\bibfield  {title} {\enquote {\bibinfo {title}
  {Quantum transport in semiconductor-superconductor microjunctions},}\ }\Doi
  {10.1103/PhysRevB.46.12841} {\bibfield  {journal} {\bibinfo  {journal} {Phys.
  Rev. B},\ }\textbf {\bibinfo {volume} {46}},\ \bibinfo {pages} {12841--12844}
  (\bibinfo {year} {1992})}\BibitemShut {NoStop}%
\bibitem [{\citenamefont {Osca}\ and\ \citenamefont
  {Serra}(2015){\natexlab{b}}}]{Os15b}%
  \BibitemOpen
  \bibfield  {author} {\bibinfo {author} {\bibfnamefont {Javier}\ \bibnamefont
  {Osca}}\ and\ \bibinfo {author} {\bibfnamefont {Lloren\c{c}}\ \bibnamefont
  {Serra}},\ }\bibfield  {title} {\enquote {\bibinfo {title} {Quasi-particle
  current in planar {Majorana} nanowires},}\ }\href
  {http://stacks.iop.org/1742-6596/647/i=1/a=012063} {\bibfield  {journal}
  {\bibinfo  {journal} {Journal of Physics: Conference Series},\ }\textbf
  {\bibinfo {volume} {647}},\ \bibinfo {pages} {012063} (\bibinfo {year}
  {2015}{\natexlab{b}})}\BibitemShut {NoStop}%
\bibitem [{\citenamefont {Groth}\ \emph {et~al.}(2014)\citenamefont {Groth},
  \citenamefont {Wimmer}, \citenamefont {Akhmerov},\ and\ \citenamefont
  {Waintal}}]{Gro14}%
  \BibitemOpen
  \bibfield  {author} {\bibinfo {author} {\bibfnamefont {Christoph~W}\
  \bibnamefont {Groth}}, \bibinfo {author} {\bibfnamefont {Michael}\
  \bibnamefont {Wimmer}}, \bibinfo {author} {\bibfnamefont {Anton~R}\
  \bibnamefont {Akhmerov}}, \ and\ \bibinfo {author} {\bibfnamefont {Xavier}\
  \bibnamefont {Waintal}},\ }\bibfield  {title} {\enquote {\bibinfo {title}
  {Kwant: a software package for quantum transport},}\ }\href
  {http://stacks.iop.org/1367-2630/16/i=6/a=063065} {\bibfield  {journal}
  {\bibinfo  {journal} {New Journal of Physics},\ }\textbf {\bibinfo {volume}
  {16}},\ \bibinfo {pages} {063065} (\bibinfo {year} {2014})}\BibitemShut
  {NoStop}%
\bibitem [{mat()}]{mathq}%
  \BibitemOpen
  \href@noop {} {}\bibinfo {note} {MATHQ code, http://www.icmm.csic.es~/sanjose
  /MathQ /MathQ. html.}\BibitemShut {Stop}%
\bibitem [{\citenamefont {Serra}(2013)}]{Serra}%
  \BibitemOpen
  \bibfield  {author} {\bibinfo {author} {\bibfnamefont {L.}~\bibnamefont
  {Serra}},\ }\bibfield  {title} {\enquote {\bibinfo {title} {Majorana modes
  and complex band structure of quantum wires},}\ }\href@noop {} {\bibfield
  {journal} {\bibinfo  {journal} {Phys.\ Rev.\ B},\ }\textbf {\bibinfo {volume}
  {87}},\ \bibinfo {pages} {075440} (\bibinfo {year} {2013})}\BibitemShut
  {NoStop}%
\bibitem [{Har()}]{Harwell}%
  \BibitemOpen
  \href@noop {} {}\bibinfo {note} {"HSL (2013). A collection of Fortran codes
  for large scale scientific computation. http://www.hsl.rl.ac.uk"}\BibitemShut
  {NoStop}%
\bibitem [{\citenamefont {Tisseur}\ and\ \citenamefont
  {Meerbergen}(2001)}]{tis01}%
  \BibitemOpen
  \bibfield  {author} {\bibinfo {author} {\bibfnamefont {Fran\c{c}oise}\
  \bibnamefont {Tisseur}}\ and\ \bibinfo {author} {\bibfnamefont {Karl}\
  \bibnamefont {Meerbergen}},\ }\bibfield  {title} {\enquote {\bibinfo {title}
  {The quadratic eigenvalue problem},}\ }\Doi {10.1137/S0036144500381988}
  {\bibfield  {journal} {\bibinfo  {journal} {SIAM Review},\ }\textbf {\bibinfo
  {volume} {43}},\ \bibinfo {pages} {235--286} (\bibinfo {year} {2001})},\
  \Eprint {http://arxiv.org/abs/http://dx.doi.org/10.1137/S0036144500381988}
  {http://dx.doi.org/10.1137/S0036144500381988} \BibitemShut {NoStop}%
\bibitem [{\citenamefont {Xie}\ \emph {et~al.}(2016)\citenamefont {Xie},
  \citenamefont {Jiang},\ and\ \citenamefont {Sha}}]{xie16}%
  \BibitemOpen
  \bibfield  {author} {\bibinfo {author} {\bibfnamefont {Hang}\ \bibnamefont
  {Xie}}, \bibinfo {author} {\bibfnamefont {Feng}\ \bibnamefont {Jiang}}, \
  and\ \bibinfo {author} {\bibfnamefont {Wei~E.I.}\ \bibnamefont {Sha}},\
  }\bibfield  {title} {\enquote {\bibinfo {title} {Numerical methods for
  spin-dependent transport calculations and spin bound states analysis in
  {Rashba} waveguides},}\ }\Doi {http://dx.doi.org/10.1016/j.cpc.2015.09.008}
  {\bibfield  {journal} {\bibinfo  {journal} {Computer Physics
  Communications},\ }\textbf {\bibinfo {volume} {198}},\ \bibinfo {pages} {118
  -- 127} (\bibinfo {year} {2016})},\ ISSN \bibinfo {issn}
  {0010-4655}\BibitemShut {NoStop}%
\bibitem [{\citenamefont {Lehoucq}\ \emph {et~al.}(1998)\citenamefont
  {Lehoucq}, \citenamefont {Sorensen},\ and\ \citenamefont {Yang}}]{arpack}%
  \BibitemOpen
  \bibfield  {author} {\bibinfo {author} {\bibfnamefont {R.~B.}\ \bibnamefont
  {Lehoucq}}, \bibinfo {author} {\bibfnamefont {D.~C.}\ \bibnamefont
  {Sorensen}}, \ and\ \bibinfo {author} {\bibfnamefont {C.}~\bibnamefont
  {Yang}},\ }\href@noop {} {\emph {\bibinfo {title} {ARPACK Users Guide:
  Solution of Large-Scale Eigenvalue Problems with Implicitly Restarted Arnoldi
  Methods}}}\ (\bibinfo  {publisher} {Philadelphia: SIAM. ISBN
  978-0-89871-407-4},\ \bibinfo {year} {1998})\BibitemShut {NoStop}%
\bibitem [{not()}]{note}%
  \BibitemOpen
  \href@noop {} {}\bibinfo {note} {In 2D the critical field and projection
  rules are not exactly given by the 1D expressions mentioned in Sec.\ I. An
  effective potential $\tilde\mu$ replaces $\mu$ in the critical field rule
  while the projection rule has small but non trivial deviations from the
  corresponding 1D expression\cite{sanjose,Os15}}\BibitemShut {NoStop}%
\end{thebibliography}%

\end{document}